\documentclass[a4paper,11pt]{article}
\pdfoutput=1 % if your are submitting a pdflatex (i.e. if you have
\usepackage{jcappub} % for details on the use of the package, please

\usepackage[T1]{fontenc} % if needed
\usepackage{verbatim}
\usepackage{csquotes}
\usepackage[style=nature,backend=biber, sorting = none, doi=false, maxbibnames=8, minbibnames=3]{biblatex}

\usepackage[dvipsnames]{xcolor}

\usepackage[english]{babel}

\usepackage{xcolor}

\usepackage{newtxtext,newtxmath}
\usepackage{ae,aecompl}
\usepackage{enumerate}
\usepackage[switch, modulo]{lineno}
\usepackage[toc,page]{appendix}
\usepackage{soul}
\usepackage{amssymb}

\usepackage{newtxtext,newtxmath}

\DeclareRobustCommand{\VAN}[3]{#2}
\let\VANthebibliography\thebibliography
\def\thebibliography{\DeclareRobustCommand{\VAN}[3]{##3}\VANthebibliography}

\usepackage{graphicx}	% Including figure files
\usepackage{amsmath}	% Advanced maths commands
\usepackage{float}

\usepackage{caption}
\usepackage{subcaption}
\usepackage{blindtext}
\title{Revisiting the CMB homogeneity scale: low multipoles removal effect and extragalactic foreground masking}

\author[a]{Xiaoyun Shao}
\emailAdd{xiaoyun48@on.br} 
\author[b]{Facundo Toscano}
\emailAdd{facundo.toscano@mi.unc.edu.ar} 
\author[b,c,d]{Diego Garcia Lambas}
\emailAdd{diego.garcia.lambas@unc.edu.ar} 
\author[a,e]{Rodrigo S. Gon\c{c}alves}
\emailAdd{rsousa@on.br} 
\author[a]{Carlos A. P. Bengaly}
\emailAdd{carlosbengaly@on.br}  
\author[b]{Heliana E. Luparello}
\emailAdd{heliana.luparello@mi.unc.edu.ar} 
\author[f]{Frode K. Hansen}
\emailAdd{frodekh@astro.uio.no}
\author[a]{Jailson Alcaniz}
\emailAdd{alcaniz@on.br}

\affiliation[a]{Observat\'orio Nacional, 20921-400, Rio de Janeiro, RJ, Brasil}
\affiliation[b]{Instituto de Astronomia Te\'orica y Experimental (IATE), CONICET-UNC, C\'ordoba, Argentina}
\affiliation[c]{Observatorio Astron\'omico de C\'ordoba (OAC), UNC, C\'ordoba, Argentina}
\affiliation[d]{Comisi\'on Nacional de Actividades Espaciales (CONAE), C\'ordoba, Argentina}
\affiliation[e]{Departamento de F\'isica, Universidade Federal Rural do Rio de Janeiro, 23897-000, Serop\'edica, RJ,  Brasil}
\affiliation[f]{Institute of Theoretical Astrophysics, University of Oslo, PO Box 1029 Blindern, 0315 Oslo, Norway}

\abstract{The Cosmic Microwave Background (CMB) reaches homogeneity  at relatively modest angular scales compared to the expectation of the standard $\Lambda$CDM model revealing an important challenge to the theoretical predictions. We analyze this inconsistency through the homogeneity scale $H$ and the slope of the homogeneity index at $\theta = 90^\circ$. We find  that the removal of low multipoles, in particular the quadrupole, from both the data and the $\Lambda$CDM synthetic CMB maps, significantly improve the consistency between models and observations. This adds to indications of the relevant  contribution of the low value of the CMB quadrupole to the observed anomalies in the homogeneity scale. Due to the presence of a new extragalactic foreground in the CMB maps, we have performed statistical analyses with different masking taking into account the regions mostly affected. In particular we consider galaxies in the local neighborhood which are expected to affect more significantly the large angular scales. We find that by masking these regions, the analysis cannot solve the discrepancy between the observations and the $\Lambda$CDM model in spite of a small improvement of their mutual consistency. The studies with both foreground masking and low-$\ell$ removed CMB maps show similar results than those of the full CMB map indicating that the main discrepancy between theory and observations is associated to the quadrupole anomaly and may require more exhaustive analysis.}

\begin{document} 
\maketitle
\flushbottom

%%%%%%%%%%%%%%%%%%%%%%%%%%%%%%%%%%%%%%%%%%%%%%%%%%
\section{Introduction}
Our standard cosmological model, known as $\Lambda$ Cold Dark Matter ($\Lambda$CDM) model, describes a Universe undergoing an accelerated expansion. Despite its success in explaining key cosmological observations, including the Cosmic Microwave Background (CMB), Large-Scale Structure (LSS) clustering via weak lensing and the luminosity distances of standard candles like Type Ia supernovae (SN)~\cite{aghanim2021planck,Brout:2022vxf,eBOSS:2020yzd,DES:2021wwk,ACT:2023kun,Li:2023tui}, the model faces both theoretical and observational challenges. \\
Theoretical issues, such as the coincidence and fine-tuning problems~\cite{weinberg2001cosmological}, question the naturalness of the $\Lambda$CDM framework. Observationally, a significant $\sim 5 \sigma$ tension exists between early-time and late-time measurements of the Hubble constant $H_{0}$ ~\cite[and references therein]{DiValentino:2021izs}. Furthermore, recent results from the first data release of the Dark Energy Spectroscopic Instrument (DESI) have hinted at possible evidence for dynamical dark energy rather than a cosmological constant~\cite{adame2025desi,calderon2024desi}. However, these findings remain under debate~\cite{cortes2024interpreting,karim2025desi,lodha2025extended,Sousa-Neto:2025gpj,mukherjee2024model}. In light of these challenges, it is essential to test the assumptions made by the $\Lambda$CDM model. One key hypothesis is the Cosmological Principle (CP), which asserts that the Universe is, from the statistical point of view, spatially homogeneous and isotropic at sufficiently large scales~\cite{clarkson2010inhomogeneity,Maartens:2011yx,Clarkson:2012bg,Aluri:2023dsf}. This principle forms the foundation of the Friedmann-Lemaître-Robertson-Walker (FLRW) metric~\cite{Hawking:1973gsd,Weinberg:2008zzc}. A direct test of the CP involves the measure of the cosmic homogeneity scale, defined as the characteristic scale beyond which the Universe appears statistically indistinguishable from a random, homogeneous and isotropic distribution (accounting for Poisson noise). These tests have traditionally employed galaxy tracers such as Quasars (QSOs) and Luminous Red Galaxies (LRGs)~\cite{Hogg,scrimgeour2012wigglez,alonso2014measuring,alonso2015homogeneity,laurent201614,ntelis2017exploring,gonccalves2018cosmic,gonccalves2018measuring,bengaly2018probing,HomogeneityQuasars,gonccalves2021measuring,andrade2022angular,mittal2024cosmic,Shao:2023sxk,Shao:2024qrd}. While homogeneity scales of $100$–$200\ h^{-1}$ Mpc are often quoted, these values largely reflect the current sampling variance of galaxy surveys rather than a real homogeneity measurement. \\
Statistical isotropy has also been tested extensively using CMB data~\cite{P13isotropy,P15isotropy,2015PhRvD..92f3008A,schwarz2016cmb,2016JCAP...06..042M,P18isotropy}, however these studies typically rely both on the angular correlation function, $\omega(\theta)$, and the power spectrum multipoles, $C_{l}$, analysis which are model-dependent estimations instead of a purely geometric measurement. Extending previous studies, \cite{Copi2009} noted the lack of correlation at angular scales $\theta \geq 60^{\circ}$, a finding that appears to challenge $\Lambda$CDM predictions. However, large uncertainties in $\omega(\theta)$ due to the observed low-$\ell$ lack of power in the CMB affect the interpretation, not being able to discard our current model~\cite{Efstathiou2010}.\\
On the other hand, recent studies have reported the existence of a new extragalactic foreground in the CMB \cite[hereafter L2023]{Luparello2023} \cite{Cruz2024}. This extragalactic foreground is detected at a level above $3\sigma$ \cite[hereafter H2025]{Hansen2024} and could explain several CMB anomalies \cite[hereafter H2023]{Hansen2023}\cite{Lambas2023}, however, without showing a significant variation of the best--fitting cosmological parameters \cite{Toscano2024}. This foreground effect seems to be strongly related to large and bright spiral galaxies and, in particular, to those in massive and dense filaments at near redshift (this is now also confirmed at the $> 3 \sigma$ level for $0.02 < z < 0.04$ in \cite{feldman2025cosmic}), showing a lower-than-average CMB temperature in these regions. Furthermore, the effect extends well beyond the galaxy halo up to $\sim 4$ $h^{-1}$Mpc in projection, indicating that the phenomenon is associated with large--scale structures with bright spirals as suitable tracers. In addition, CMB photons passing through local voids ($z < 0.03$) have been found to be hotter than expected at the $2.7-3.6 \sigma$ level \cite{hansen2025evidence}.\\
Taking into account both, the lack of power in the observed low-$\ell$ spectrum and the new extragalactic foreground that seems to affect the CMB observations, we based our work on \cite[CQG2022 hereafter]{camacho2022measurement} where the authors applied a pure geometrical fractal dimension ($H(\theta)$) analysis to the Early Universe, offering a model-independent approach to study the homogeneity. Unlike $\omega(\theta)$ or $C_{l}$, the fractal dimension $H(\theta)$ depends on the scaling behavior of fluctuations rather than their amplitude. \\
Due to that, initially we study the effect of low multipoles on the angular correlation function $w(\theta)$ in order to understand at what limits these multipoles can cause the discrepancy between the model and the observations. Since the CMB extragalactic foregrounds act on the low-$\ell$s, this measurement can explain the impact of them on the homogeneity scale. After that, at the same fashion as CQG2022 applied in their work, we study the CMB homogeneity scale through the analysis of $H(\theta)$, firstly reproducing their results and then investigating how the foregrounds affect the estimations.

The manuscript is structured as follows: In Section~\ref{sec:DATA_ANALYSIS}, we describe our data sample and the respective theoretical framework, where we explore the statistical analyses and estimators. In section~\ref{sec:RESULTS_DISCUSSION}, we present and discuss our main results. Finally, we summarize our conclusions in Section~\ref{sec:CONCLUSIONS}.

\section{Data and Analysis}
\label{sec:DATA_ANALYSIS}

\subsection{CMB Data}
$ $

\begin{figure}
    \centering
    \includegraphics[width=0.5\linewidth]{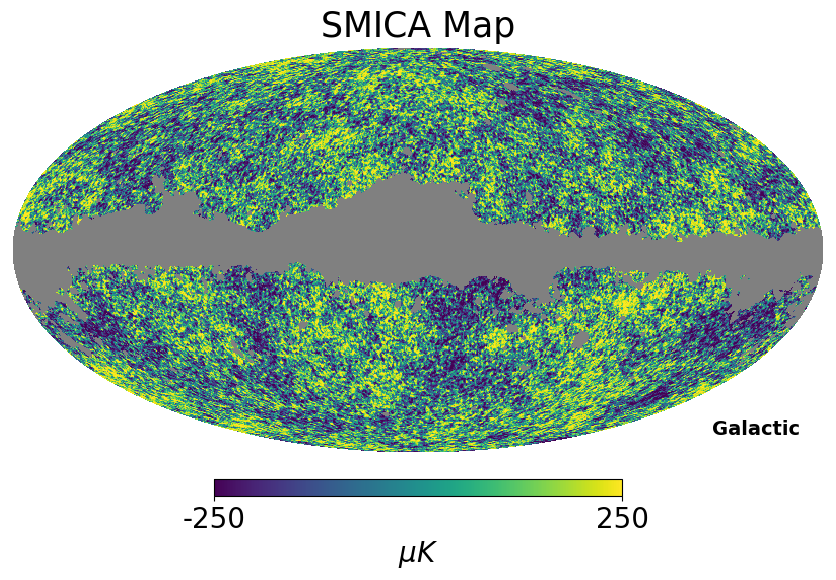}
    \caption{Planck SMICA map with the common mask in galactic coordinates.}
    \label{fig:SMICA}
\end{figure}

In order to explore the statistical homogeneity of the CMB, we need to choose the appropriate maps and masks for the analysis. We use the SMICA maps from Planck Public Release 3 (Planck PR3)\footnote{Data available in \url{https://pla.esac.esa.int/}}. To mask foreground residuals around the Galactic plane as well as bright extragalactic point sources, we apply the Planck common mask based on the union of the uncertainties of four different subtraction methods used in the Planck PR3 analysis \cite{akrami2018planck,akrami2020planck}, leaving about $78\%$ of the sky area suitable for statistical studies. For the purpose of measuring the statistical significance of the sample, we generate 1000 synthetic maps based on the best-fitting $\Lambda$CDM spectrum given by Planck PR3 \cite{P18isotropy}. On the other hand, we use a jackknife method to measure the intrinsic sample variance of our data, dividing the sky into $N_{JK} = 128$ approximately equal regions. In Figure \ref{fig:SMICA} it can be seen the SMICA CMB map with the common mask in galactic coordinates.

\subsection{CMB Statistical Estimators}
In this section, we describe the theoretical framework used to quantify both effects previously mentioned: the low-$\ell$ multipoles on the temperature angular correlation function and the foregrounds on the homogeneity scale. For the former we review the temperature two-point angular correlation function (2PACF) on the CMB and for the latter we detail the standard approach of the fractal dimension $H(\theta)$ employed in galaxy surveys in order to demonstrate its adaptation to the Cosmic Microwave Background.

\subsubsection{CMB Two--Point Angular Correlation Function}
As it is explained in \cite[and references therein]{Copi2015}, the CMB Temperature 2PACF computed as an average over the complete sky was found to be smaller than expected at large angular scales ($\theta > 60^{\circ}$). Notice that comoving scales that cross into the Hubble radius at $z \sim 1$ and below are observed at angles larger than $60^{\circ}$, relating the features observed at those scales to either a primordial nature or belonging to astrophysical processes corresponding to structures at that redshift. These angular scales are roughly associated to the scales that have shown an anomalous behavior with respect to the predictions, named the quadrupole, octopole, and up to modes $\ell = 10$. Moreover, the large-angle CMB is also sensitive to the physics that affects the microwave photons as they propagate from the last scattering until us. Indeed, the Integrated Sachs–Wolfe (ISW) effect could potentially correlate the large-angle CMB with the local structure of
the gravitational potential. \\
For the analysis of the full-sky CMB map, $T(\hat{e})$ can be expanded in spherical harmonics as
\begin{equation}
    T(\hat{e}) = \sum_{\ell, m} a_{\ell m}Y_{\ell m}(\hat{e}),
\end{equation}
where
\begin{equation}
    a_{\ell m} = \int T(\hat{e}) Y_{\ell m }^* (\hat{e}) d\hat{e}.
\end{equation}
From these expressions, we can define the angular power spectrum as
\begin{equation}
    C_{\ell} = \frac{1}{2\ell +1} \sum_m |a_{\ell m}|^2.
\end{equation}
One can associate the full-sky angular power spectrum with the two-point angular correlation function via
\begin{equation} \label{C_theta}
    C(\theta) = \sum_{\ell} \frac{2\ell +1}{4 \pi} C_{\ell}P_{\ell}(cos \theta)
\end{equation}
where the $P_{\ell}(cos\theta)$ are Legendre polynomials. Because of the foreground contamination, a full-sky CMB map is not available. If the mask on the sky is represented by $W(\hat{e})$, then the cut-sky 2PACF is defined as:
\begin{equation} \label{2PACF}
 C^{cut}(\theta) = \overline{W(\hat{e}_1)T(\hat{e}_1)W(\hat{e}_2)T(\hat{e}_2)}.
\end{equation}
Defining the pseudo-$a_{\ell m}$ as
\begin{equation}
    \tilde{a}_{\ell m} = \int W(\hat{e})T(\hat{e})Y^*_{\ell m}(\hat{e})d\hat{e},
\end{equation}
the pseudo-$C_{\ell}$ can be written like
\begin{equation} \label{power_spectrum_cut}
    \tilde{C}_{\ell} = \frac{1}{2\ell +1} \sum_m |\tilde{a}_{\ell m}|^2.
\end{equation}
An effective method to study the effect of the low-$\ell$ multipoles on the 2PACF is the cutting of the lowest ones ($\ell \leq 2$ or $\ell \leq 5$) on the angular power spectrum (Eq. \ref{power_spectrum_cut}), obtaining a 2PACF (Eq. \ref{2PACF}) that does not take into account the contribution of these angular scales.
\subsubsection{Homogeneity Scale Index} \label{sec:H}
Following the steps in CQG2022, we can quantify the homogeneity through the homogeneity scale index $H(\theta)$. This estimator is derived from the scaled counts-in-spheres $\mathcal{N}(<\theta)$, where specifically $H(\theta)$ is the logarithmic derivative of the scaled counts-in-spheres:
\begin{equation} \label{h_index}
    H = 2 \frac{d\ln{\cal{N}(<\theta)}}{d\ln{\Omega}}
\end{equation}
The homogeneity scale $\theta_{H}$ is defined as the angular scale where $H =0$, i.e.
\begin{equation}
\theta_{H} \equiv \theta(H=0)\;. 
\label{thetah8}
\end{equation}
Considering the definition of the CMB Temperature 2PACF (Eq. \ref{C_theta}), we can adapt the homogeneity index for the CMB case through: 
\begin{equation}
W_\eta (\theta) \equiv 1 +  C(\theta)\;.
    \label{Eq:w-t}
\end{equation}
A characteristic scale can be defined at the crossing point where $C(\theta)=0$, so $W_\eta=1$,
\begin{equation}
\theta_{W} \equiv \theta(W_\eta=1)\;.
\label{thetah8}
\end{equation}
The homogeneity index, originally defined in Eq.~\ref{h_index}, is rewritten as
\begin{equation}
    H(\theta) = 2\frac{d\ln{\mathcal{P}(<\theta)}}{d\ln{\Omega(<\theta)}}\;,
    \label{Eq:homogeneity-index-h}
\end{equation}
where \begin{equation}
    \mathcal{P}(<\theta) = \frac{P<\theta)}{\tilde{P}(<\theta)} = \frac{1}{1-\cos{\theta}} \int_0^{\theta} W_\eta(\theta) \sin{\theta}d\theta\;.
\label{Eq:homogeneity-index-T}
\end{equation}
To estimate statistical uncertainties, the map is divided into $N_{JK}$ Jackknife regions. The mean correlation function and its variance are computed as:
\begin{equation}
    \Bar{W}_\eta(\theta) = \frac{1}{N_{JK}}\sum_i W_{\eta i}(\theta),
    \label{Eq:mean-w}
\end{equation}
and  the error of the Jackknife regions are estimated as
\begin{equation}
\sigma^2= \frac{N_{JK}-1}{N_{JK}}\sum_i\left[ W_{\eta i}(\theta)-\Bar{W}_\eta(\theta) \right]^2 ,
\label{Eq:error-w}
\end{equation}
For a more detailed study of the homogeneity scale, we introduce another useful measurement based on the slope of the homogeneity index at $\theta=90^{\circ}$, defined as
\begin{equation}
    S_{90}=\left.\frac{d H(\theta)}{d \theta}\right|_{\theta =90^{\circ}}\;.
    \label{s90}
\end{equation}
which also characterizes the global behavior of the CMB fluctuations.

\begin{figure*}
    \centering \includegraphics[width=0.49\linewidth]{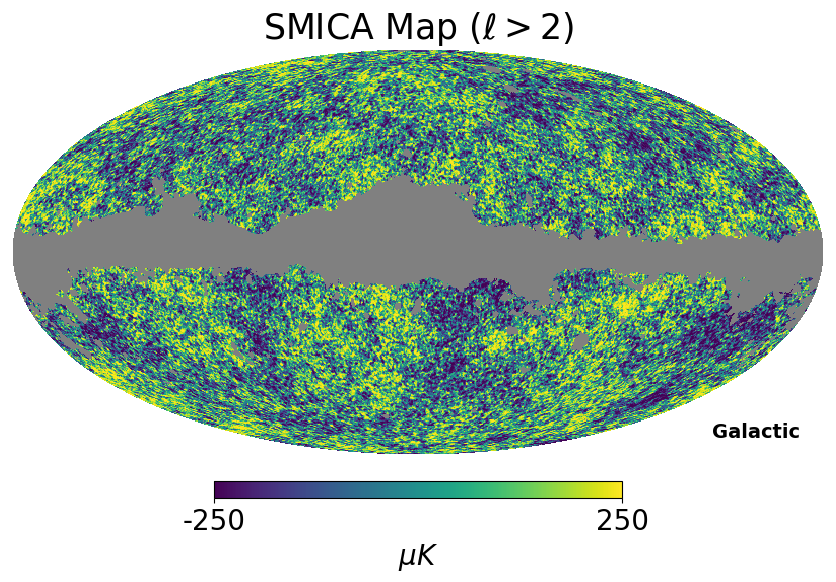}
\includegraphics[width=0.49\linewidth]{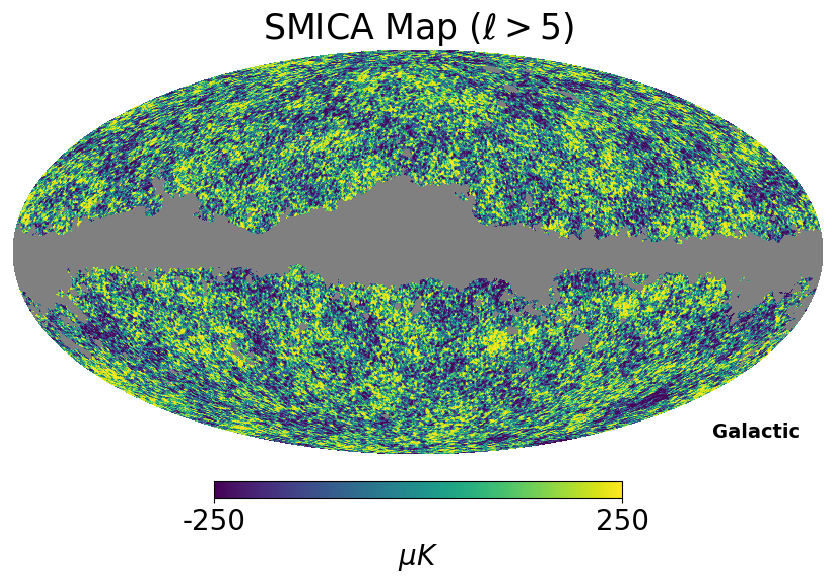}
    \caption{
    Planck SMICA map with the removal of the first 3 multipoles (left) and the first 6 multipoles (right).
    }
    \label{removed_maps}
\end{figure*}

\subsection{Low $\ell$ remotion and Foreground masking maps} \label{sec:FOREGROUND_MODELS}
$ $

As shown in L2023 and H2025, the extragalactic foreground effect is statistically significant for the low CMB multipoles ($\ell < 100$), being related to the large--scale structures surrounding large and bright spiral galaxies in the nearby universe. On the other hand, it is well known the lack of power for the CMB low-$\ell$ multipoles, where the observed first multipoles signal is much smaller than the predicted ones. Therefore, we remove the first multipoles of both, the observed CMB data and synthetic maps in order to assess the relevance of these low $\ell$ multipoles in the agreement between the observations and the theoretical $\Lambda$CDM predictions. This allows to estimate to what extent the presence of foreground effects on large angular scales can affect the correlation function $W_\eta$, and the homogeneity index $H$.\\
For this purpose, we construct two maps with \texttt{NSIDE = 256} removing the first three and six multipoles, respectively, as shown in Figure \ref{removed_maps}. Together with these maps, we generate 1000 synthetic maps with the same characteristics in order to obtain a suitable cosmic variance estimation. These simulations are produced using the \texttt{synfast} routine of the Healpix package\footnote{\url{https://healpix.sourceforge.io/}} \cite{gorski2005healpix}.%\footnote{\url{http://healpix.jpl.nasa.gov/}}.
We follow the methodology of H2025 in order to reduce correlations with the smaller scales in the estimation of the new maps. The smallest multipoles are removed using an extension of the \texttt{remove\_dipole} procedure of the Healpix package to multipoles beyond $\ell=1$. As explained in H2025, in order to avoid aliasing of smaller scales into the larger scales, we degrade the original map in order to estimate the smallest multipole moments. We proceed to downgrade the resolution of the CMB map to $N_\mathrm{side}=4$. We apply the common mask before degrading, and then we extend it correspondingly. At this resolution, multipoles up to $\ell_\mathrm{max}=4\times N_\mathrm{side}=16$ are maintained. On this downgraded map, we use a $\chi^2$ minimization approach to estimate all the individual $\hat a_{\ell m}$ from $\ell=0$ to $\ell=16$:
\begin{equation}
    \chi^2 = \sum {\bf d}^\dagger C^{-1}{\bf d} \ ,
\end{equation}
where the elements of the data vector are given as
\begin{equation}
    d_{\ell m} = \sum_{\Omega}\hat a_{\ell m} Y_{\ell m}(\Omega) M(\Omega)- \tilde a_{\ell m}
\end{equation}
and the sum is performed over all the pixels of the map and $\tilde a_{\ell m}$ are the pseudo-$a_{\ell m}$ with the extended mask $M(\Omega)$. The coupling matrix is calculated as
\begin{equation}
    C_{\ell m,\ell' m'}=\sum_{\Omega}M(\Omega)Y_{\ell m}(\Omega)Y_{\ell' m'}(\Omega) \ .
\end{equation}
From the estimated $\hat a_{\ell m}$, we create multipole maps for each $\ell$ as
\begin{equation}
  \hat M_\ell(\Omega) = \sum_{m=-\ell}^\ell \hat a_{\ell m}Y_{\ell m}(\Omega) \ .
\end{equation}
In this way, we obtain $M_\ell(\Omega)$ for the CMB map up to $\ell=16$. We ensure that by summing all estimated multipole maps up to $\ell=16$, the original $N_\mathrm{side}=4$ map is obtained to numerical precision. Repeating the procedure on simulated maps with known full-sky multipoles, we find a maximum $9\mu$K error in some pixels very close to the mask, gradually decreasing away from the mask. Then we remove (1) the estimated quadrupole map and (2) all multipoles maps up to $\ell\leq5$ from the original high resolution map.

\begin{figure*}[h]
    \centering
    \includegraphics[width=0.325\textwidth]{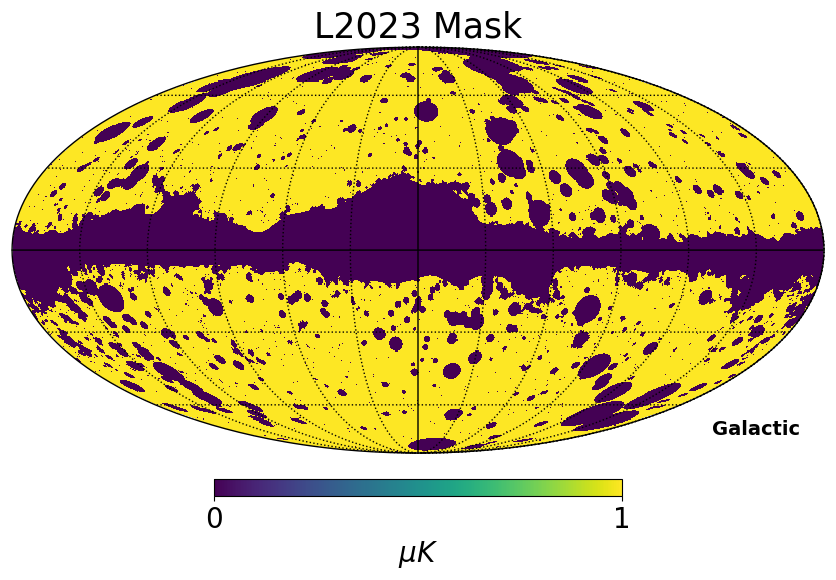}
    \includegraphics[width=0.325\textwidth]{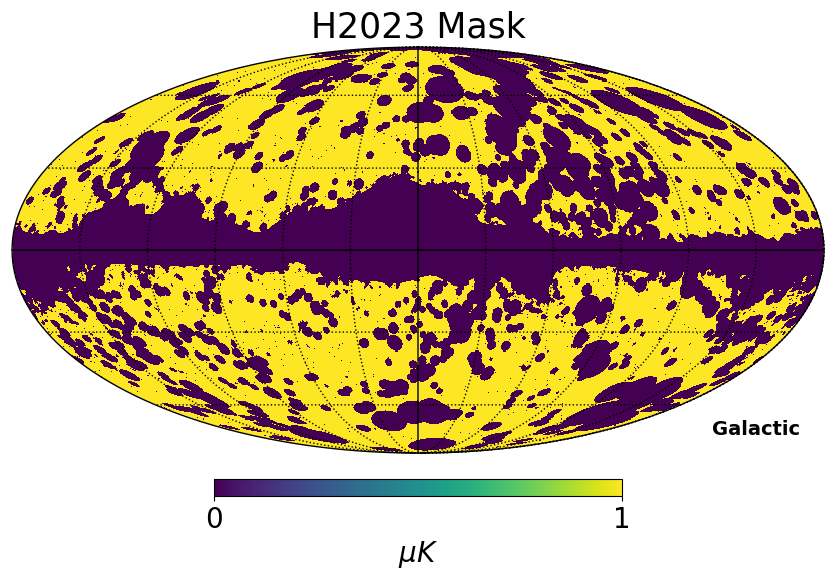}
    \includegraphics[width=0.325\textwidth]{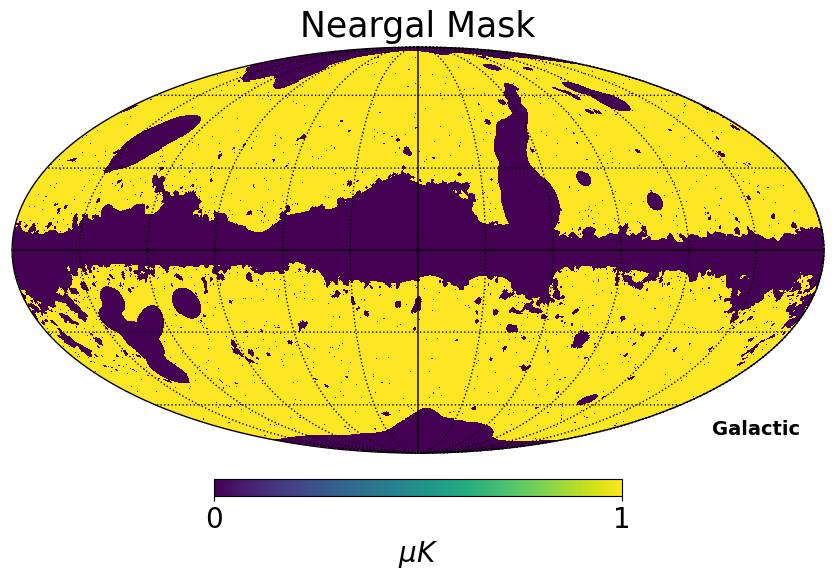}
    \caption{Comparison among the L2023 (left), H2023 (middle), and Neargal (right) foreground masks.}
    \label{all_mask}
\end{figure*}

Besides the previous study we have also considered the possible effect of the extragalactic foreground in the correlation function and the homogeneity index through masking the regions where the foreground is expected to be more significant. For this aim we have used L2023, H2023 and a new masked region corresponding to the most nearby galaxies which correspond to large angular regions in the sky and that were excluded from the L2023 and H2023 studies due to the effects of peculiar velocities. \\
The L2023 foreground mask is related to the distribution of large (K-band image radius greater than 8.5 kpc) spiral galaxies in the redshift range $z= [0.001, 0.017]$. The H2023 foreground mask corresponds to the model of H2023 extended to galaxies at higher redshifts. The foreground mask of the most nearby galaxies is based on galaxy positions and redshift independent distances from the Nearby Galaxy Catalogue \cite{Karachentsev2013} with  radius $r> 7.5$ kpc (Neargal Mask). A total of 78 regions were masked corresponding to circles of 2 degree radius. For this mask associated to the most nearby galaxies, the remaining sky fraction is $70\%$. Figure~\ref{all_mask} shows the three foreground masks analyzed.

\section{Results and Discussion} \label{sec:RESULTS_DISCUSSION}

\subsection{Low-$\ell$ multipole removal effect}
\begin{figure*}
    \centering \includegraphics[width=0.8\textwidth]{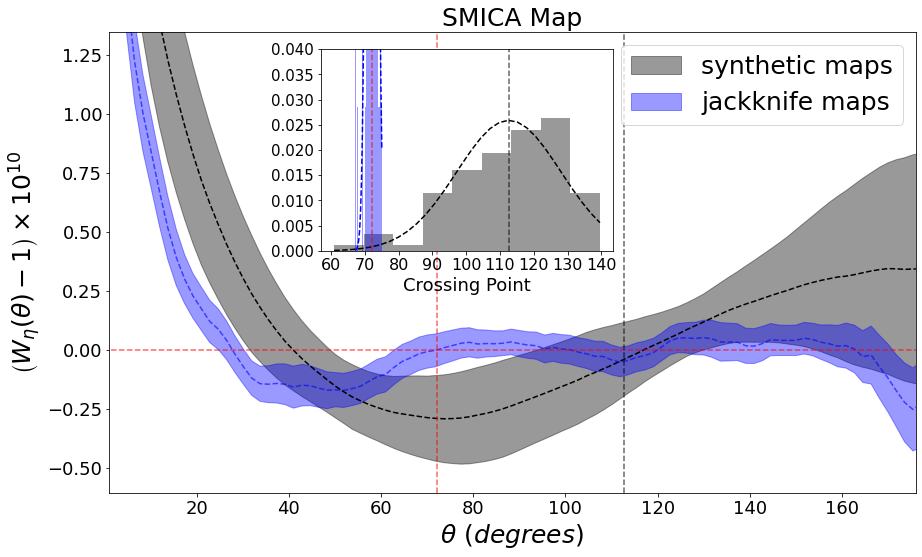}
    \caption{Angular correlation function $W_{\eta}(\theta)$ for SMICA map masked by the common mask, compared to $\Lambda$CDM synthetic maps. }
    \label{w_theta_lcdm}
\end{figure*}

\begin{figure*}
    \centering
    \begin{subfigure}{0.49\textwidth}
        \centering
        \includegraphics[width=\linewidth]{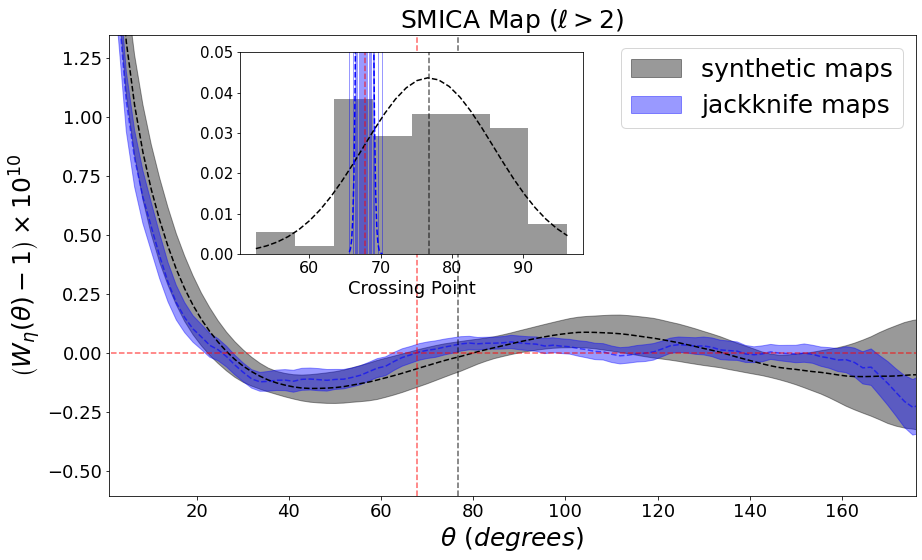}
    \end{subfigure}
    \begin{subfigure}{0.49\textwidth}
        \centering
        \includegraphics[width=\linewidth]{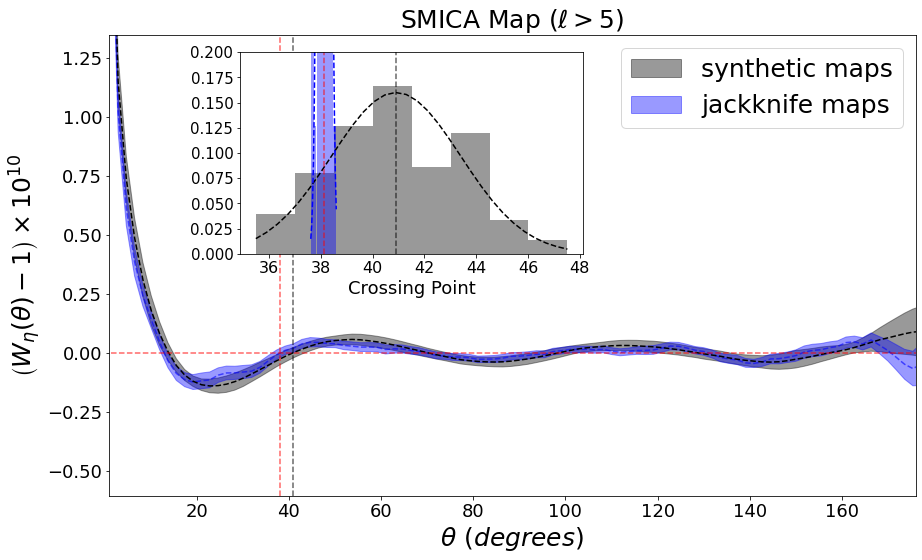}
    \end{subfigure}
    \caption{Angular correlation function $W_\eta(\theta)$ for the Planck SMICA map with low multipoles removed: $\ell > 2$ (left) and $\ell > 5$ (right). In both panels, the dashed black lines show the mean from synthetic $\Lambda$CDM realizations, with the grey shaded regions indicating their 1$\sigma$ dispersion. The blue dashed lines represent the SMICA data, while the blue shaded regions denote the  estimated  uncertainty. The insets display the distribution of the second crossing point $\theta_W$, defined by $W_\eta(\theta) = 1$, with vertical dashed lines indicating the mean crossing angles from the simulations (black) and data (red).  }
    \label{w_theta_cutted}
\end{figure*}

\begin{figure*}
	\centering	\includegraphics[width=0.8\textwidth]{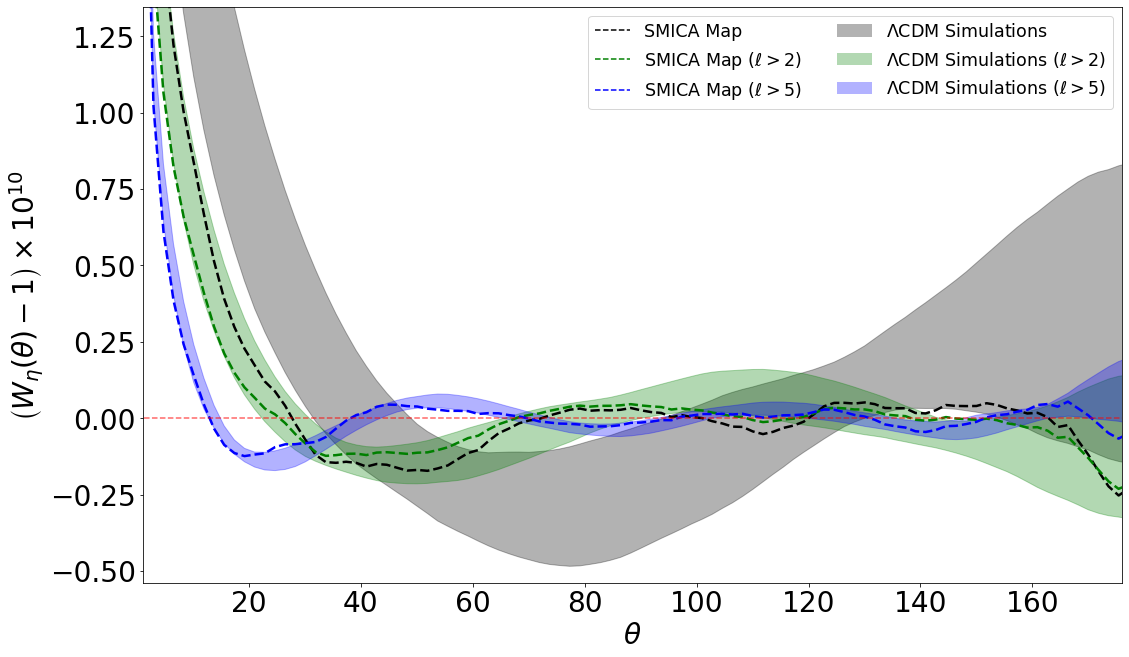}
        \caption{Comparison of $W_{\eta}(\theta)$ for SMICA maps and $\Lambda$CDM simulations with varying removal of low-$\ell$ multipoles. The black dashed line represents the original SMICA map, while the green and blue dashed lines correspond to the SMICA maps with the first three ($\ell \leq 2$) and six ($\ell \leq 5$) multipoles removed, respectively. Shaded regions show the $1\sigma$ spread from  $\Lambda$CDM simulations for each case.}
        \label{Combined_wtheta}
\end{figure*}

In Figure \ref{w_theta_lcdm} we show the correlation function for the SMICA map together with the $\Lambda$CDM prediction. The blue dashed line and shaded region represent the mean and 1$\sigma$ uncertainty derived from jackknife resampling of the observed data, while the black dashed line and gray shaded region correspond to the mean and 1$\sigma$ range from  synthetic $\Lambda$CDM realizations. The inset shows the distribution of the second crossing angle $\theta_W$, defined by $W_\eta(\theta_W) = 1$. The red vertical line marks the mean value of $\theta_W$ from the SMICA map, and the black dashed line indicates the mean from simulations. Even taking into account both data uncertainty and cosmic variance, the discrepancy between them is notable. Moreover, in Figure \ref{w_theta_cutted} we show $W_n$ for both cases, where $0 \leq \ell \leq 2$ is removed and $0 \leq \ell \leq5$ is removed. The data uncertainty (calculated through the Jackknife method) and the cosmic variance (estimated through simulations) are plotted as well. In Figure \ref{Combined_wtheta}, a comparison of these results and the theoretical $\Lambda$CDM prediction is presented. The results of the removal of the first multipoles show that even for a small number of multipoles removed (2 and 5, respectively), the discrepancy between the observations and the predictions is no longer present. This is a clear indication that the homogeneity scale tension with the $\Lambda$CDM models is mainly associated to the low quadrupole value of the CMB. In order to test the robustness of this conclusion given possible errors in the multipole removal process, we repeated the procedure, using the official Planck SMICA impainted map. Removing the quadrupole from the full-sky impainted map resulted in $W_\eta$ which are very similar to the results presented in Figure \ref{w_theta_cutted}. We therefore conclude that the exact procedure for removing the quadrupole does not affect the results.\\
    Numerically, Table~\ref{tab:t_cross_points7} summarizes the statistical properties of the estimators $\theta_{W}$, $\theta_{H}$, and $S_{90}$ after the removal of different low multipoles, along with their corresponding reduced chi-squared values using the common mask, as presented in Table~\ref{tab:t2_combined_chi_squared}. 
The chi-squared ($\chi^2$) statistic is computed following the method described in \cite{Gaztanaga2003}, given by
\begin{equation}
    \chi^{2}=\sum_{i=1}^{i=N_{b}} \sum_{j=1}^{j=N_{b}} \Delta_{i} \hat{C}_{i j}^{-1} \Delta_{j} \, ,
    \label{eq1}
\end{equation}
where the residual vector $\Delta_i$ is defined as
\begin{equation}
\Delta_{i}=\frac{W^{\text {obs }}  (i) -W^{\bmod }  (i) }{ \sigma_{W}  (i)  } \, ,
\end{equation}
Here, $W^{\text{obs}}$ denotes the observed angular correlation function, while $W^{\text{mod}}$ represents the model prediction. In our analysis, $W^{\text{mod}}$ is taken as the mean $W^{\text{syn}}$ computed from 1000 synthetic CMB maps. \\   
In Table \ref{tab:t_pvalue_all1} we present the p-values associated with the homogeneity indicators $\theta_W$, $\theta_H$, and the slope $S_{90}$, computed from the SMICA CMB maps with and without the removal of the lowest multipoles. Following the methodology of Planck Collaboration XVI (I$\&$S) and \cite{muir2018covariance}, we can calculate p-values to quantify how the observed temperature maps are unusual with respect to the $\Lambda$CDM simulations, where values below 2.2$\%$ indicate $\sim$2$\sigma$ tension and those above 16$\%$ are considered consistent with the model (i.e., <1$\sigma$). For the full SMICA map, $\theta_W$ and $\theta_H$ yield p-values of 1.0$\%$, suggesting moderate tension with $\Lambda$CDM. However, after removing the lowest multipoles ($\ell \leq 2$), the p-values increase significantly—especially for $\theta_H$ (23.0$\%$)—indicating improved agreement. The trend continues with the removal of $\ell \leq 5$, where $\theta_H$ and $S_{90}$ reach 32.0$\%$ and 33.0$\%$, respectively, well within the no-tension regime. These results support the conclusion that low-$\ell$ modes contribute substantially to observed deviations and that their exclusion enhances the consistency of homogeneity-related observables with $\Lambda$CDM expectations.
\begin{table}[!t]
    \centering
    \renewcommand{\arraystretch}{1.2}
    %\resizebox{\textwidth}{!}{  % Rescale to fit full page width
    \begin{tabular}{lcccc}
        \hline
        \textbf{ } & \textbf{SMICA} & \textbf{SMICA $(\ell > 2)$} & \textbf{SMICA $(\ell > 5)$} \\
        \hline
        $ \theta_W $ (Data)  & 72.12 ± 13.95 & 67.76 ± 6.28 & 38.13 ± 1.84 \\
        $ \theta_W $ (Sim)  & 112.64 ± 15.48 & 76.72 ± 9.18 & 40.91 ± 2.5 \\
        $ \theta_H $ (Data)  & 66.84 ± 8.78 & 63.47 ± 4.23 & 35.56 ± 1.45 \\
        $ \theta_H $ (Sim)  & 105.52 ± 14.49 & 71.07 ± 9.16 & 36.72 ± 2.22 \\
        $ S_{90} \times 10^{-12} $ (Data)  & -0.649 ± 0.949 & -0.451±0.544  & 0.37±0.406 \\
        $ S_{90} \times 10^{-12} $ (Sim)  & 2.073 ± 1.927 & 0.603± 1.023 &  0.598±0.549 \\
        \hline
    \end{tabular}
    %}
    \caption{ $\theta_W$, $\theta_H$, and $S_{90}$ after the removal of low multipoles, derived from SMICA map (Data) and $\Lambda$CDM simulations (Sim). All maps are masked using the common mask.}
    \label{tab:t_cross_points7}
\end{table}
\begin{table}[!t]
    \centering
    \renewcommand{\arraystretch}{1.2}
    %\resizebox{\textwidth}{!}{  % This rescales to fit the page
    \begin{tabular}{ccccc}
        \hline
        & \textbf{$N_{JK}$} & \textbf{SMICA} & \textbf{SMICA $(\ell > 2)$} & \textbf{SMICA $(\ell > 5)$} \\
        \hline
        $W(\theta)$ & 128 & 6.96 (167.05/24) & 1.15 (27.53/24) & 0.88 (32.41/37) \\
        $H(\theta)$ & 128 & 7.27 (159.85/22) & 1.27 (27.93/22) & 1.06 (31.93/30) \\
        $W(\theta)$ & 256 & 8.16 (252.94/31) & 1.36 (39.45/29) & 0.92 (40.48/44) \\
        $H(\theta)$ & 256 & 8.28 (223.6/27) & 1.48 (39.95/27) & 1.07 (37.4/35) \\
        \hline
    \end{tabular}
    %}
    \caption{Reduced chi-squared ($\chi^2_\nu$) values for the angular correlation function $W(\theta)$ and the homogeneity index $H(\theta)$, computed from Planck SMICA maps  for different numbers of jackknife regions ($N_{JK} = 128$ and 256).  Results are shown for the full SMICA map, as well as for maps with the first three ($\ell > 2$) and six ($\ell > 5$) multipoles removed. Each entry displays the reduced $\chi^2$ value followed by the total $\chi^2$ and degrees of freedom in parentheses.}
    \label{tab:t2_combined_chi_squared}
\end{table}
\begin{table}[!t]
    \centering
    \renewcommand{\arraystretch}{1.2}
    \begin{tabular}{lcccc}
        \hline
        \textbf{ } & \textbf{SMICA} & \textbf{SMICA $(\ell > 2)$} & \textbf{SMICA $(\ell > 5)$}  \\
        \hline
        $ \theta_W $  & 1.0 $\%$  & 18.0$\%$  & 12.0$\%$\\

        $ \theta_H $ & 1.0$\%$  & 23.0$\%$   & 32.0$\%$\\

        $ S_{90}  $  & 8.0$\%$  & 13.0$\%$  & 33.0$\%$\\

        \hline
    \end{tabular}
    \caption{P-values associated with the homogeneity scales $\theta_{W}$ and $\theta_{H}$, and the slope $S_{90}$, computed from the SMICA CMB maps. Results are shown for the full map and for maps with the lowest multipoles removed ($\ell > 2$ and $\ell > 5$), all processed using the common mask. The P-values quantify the probability of obtaining values as extreme as the observed ones under the $\Lambda$CDM model assumptions.
 }
    \label{tab:t_pvalue_all1}
\end{table}

\subsection{Foreground masking on CMB statistical estimator results}\label{sec:3.2}
\begin{figure}[ht]
    \centering
    \begin{minipage}[b]{0.49\textwidth}
        \includegraphics[width=\textwidth]{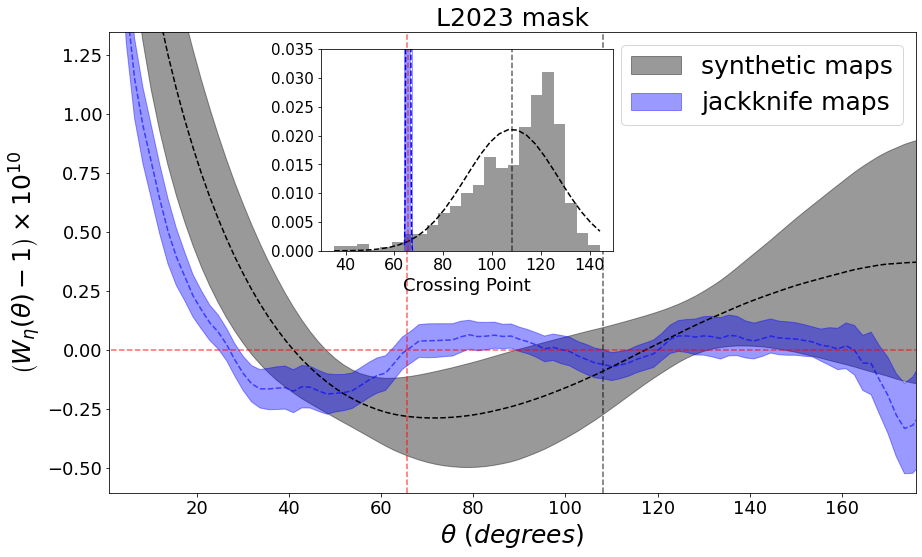}
    \end{minipage}
    \hfill
    \begin{minipage}[b]{0.49\textwidth}
        \includegraphics[width=\textwidth]{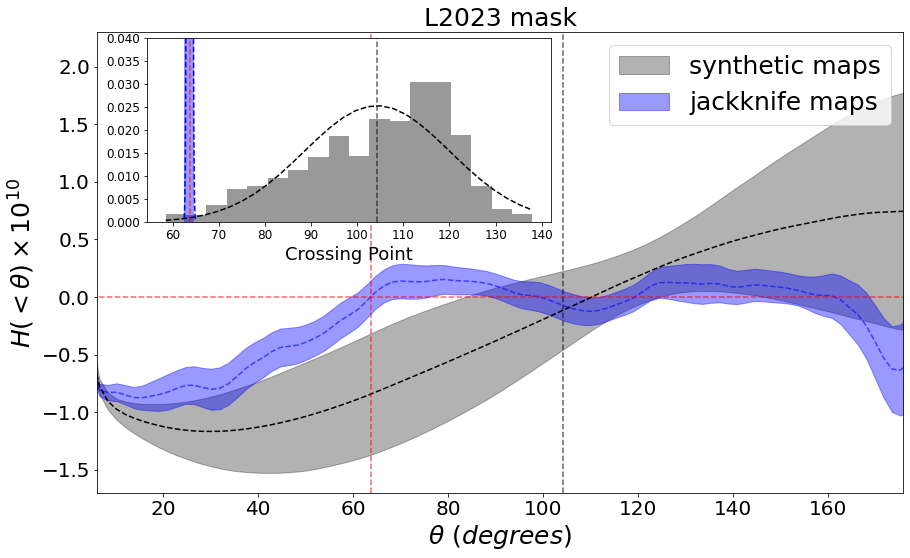}
    \end{minipage}
        \hfill
    \begin{minipage}[b]{0.49\textwidth}
        \includegraphics[width=\textwidth]{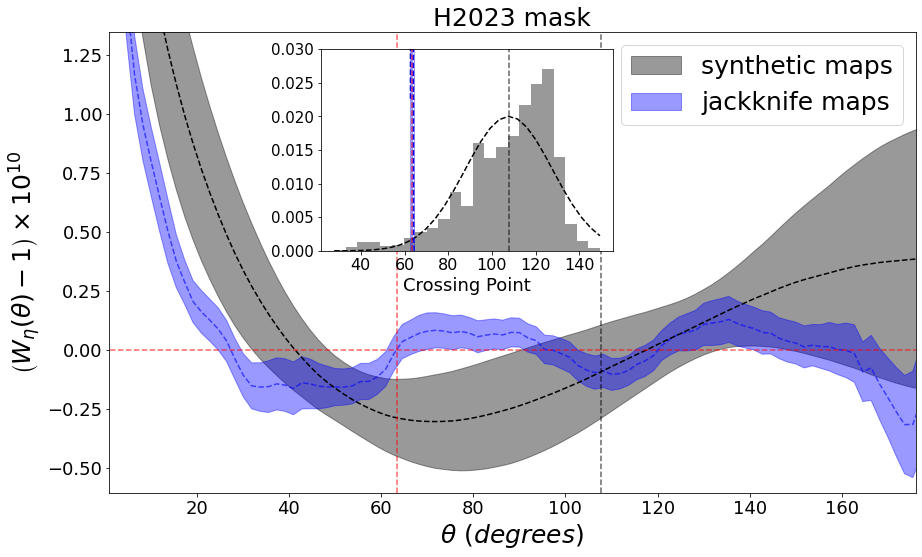}

    \end{minipage}
    \hfill
    \begin{minipage}[b]{0.49\textwidth}
   \includegraphics[width=\textwidth]{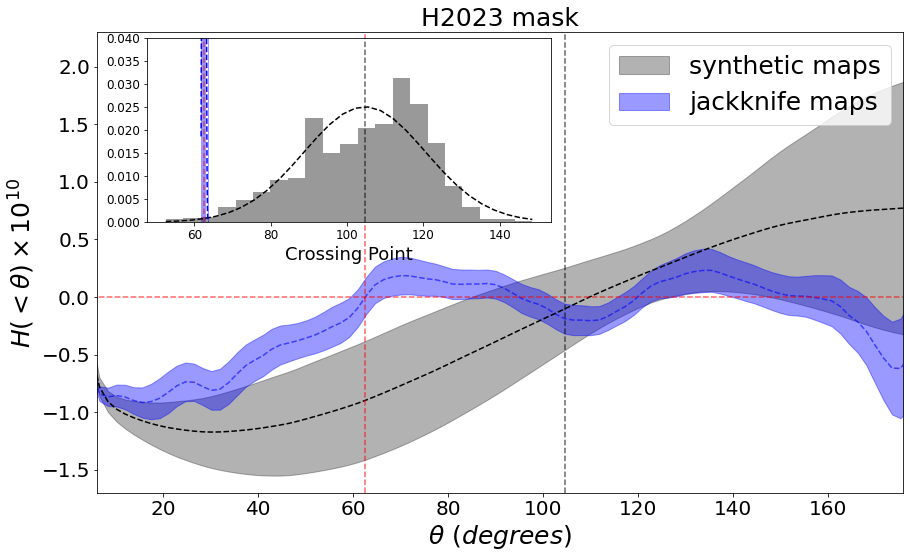}
    \end{minipage}
        \hfill
    \begin{minipage}[b]{0.49\textwidth}
        \includegraphics[width=\textwidth]{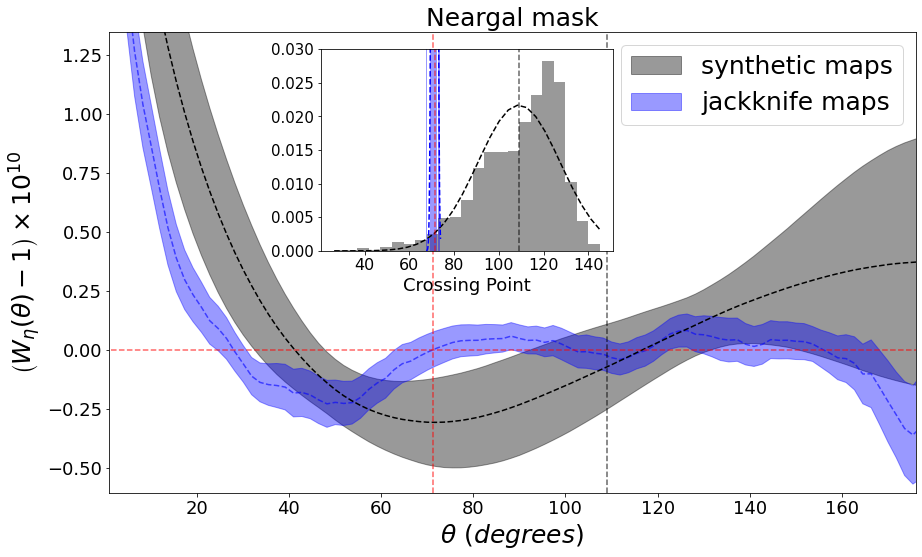}
    \end{minipage}
     \hfill
    \begin{minipage}[b]{0.49\textwidth}      \includegraphics[width=\textwidth]{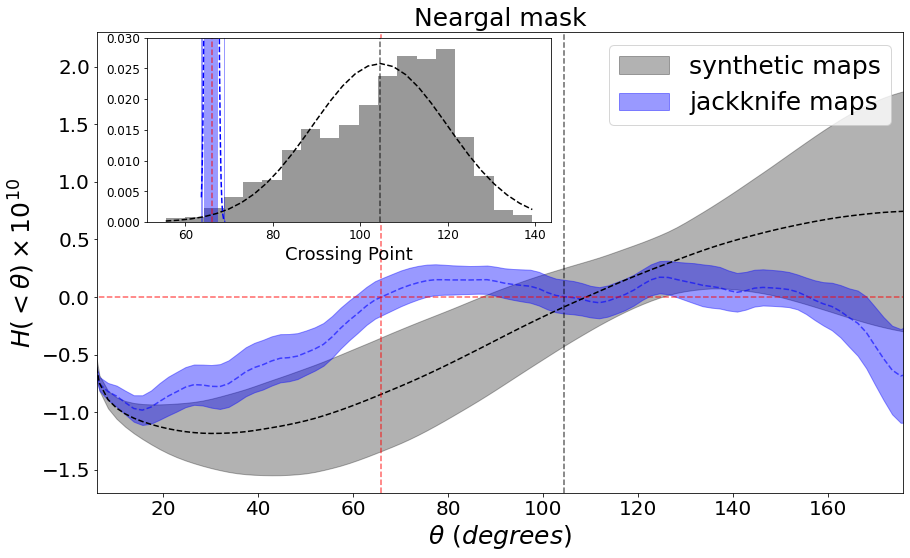}
    \end{minipage}
        \caption{The $1\sigma$ uncertainties of $W(\theta)$ (left column) and $H(\theta)$ (right column) are shown. Each row corresponds to the SMICA map masked by the L2023 mask (top), the H2023 mask (middle), and the Neargal mask (bottom).}
        \label{fig:wh9}
\end{figure}

In order to explore how foreground masking influences CMB statistical properties, we analyze the angular correlation function $W_\eta(\theta)$ and the homogeneity index $H(\theta)$. Three masking scenarios are considered: using L2023, H2023 and Neargal masks. In Fig.~\ref{fig:wh9} one can visualize the resulting statistical estimators for the three masks analyzed. Their corresponding uncertainties derived from 128 Jackknife regions and 1000 synthetic maps are also presented. The angular correlation function $W_\eta(\theta)$ is presented on the left column while the $\mathcal{H(<\theta)}$ is presented on the right. For all cases, the masked data are shown as a blue dashed line, while the results from the masked synthetic maps are shown as a black dashed line. The $1\sigma$ uncertainties from the data and simulations are represented by the shaded regions.
The accompanying histograms show the distribution of the angular scale $\theta_W$ ($\theta_{H}$), defined as the second crossing point where \( W_\eta(\theta) = 1 \) ($H(\theta) = 0$). The red dashed line marks the mean $\theta_W$ ($\theta_{H}$) from the data and the black dashed line corresponds to the mean value from the $\Lambda$CDM simulations. 
 \\
Also, we compare the statistical properties of three key estimators — $\theta_{W}$, $\theta_{H}$, and $S_{90}$ — under the three masking methods, as summarized in Table~\ref{tab:cross_points}. We evaluate the 1$\sigma$ uncertainties obtained from both Jackknife resampling and synthetic $\Lambda$CDM realizations for the masked SMICA map with the L2023, H2023 and Neargal masks. 
\begin{table}[h]
    \centering
    \renewcommand{\arraystretch}{1.2}
    \begin{tabular}{lcccc}
        \hline
        \textbf{ } & \textbf{Common Mask} & \textbf{L2023} & \textbf{H2023} & \textbf{Neargal Mask} \\
        \hline
        $ \theta_W $ (Data)  & 72.12 ± 13.95 &65.68 ± 4.74 &63.37 ± 2.6 & 71.27 ± 8.95    \\
        $ \theta_W $ (Sim)  & 112.64 ± 15.48 &108.21±10.89&107.73±19.94& 106.15 ± 21.04  \\
        $ \theta_H $ (Data)  & 66.84 ± 8.78   & 63.71±3.55 & 62.56±2.87 & 65.97 ± 8.34     \\
        $ \theta_H $ (Sim)  & 105.52 ± 14.49 & 104.3 ± 15.81 & 104.69 ± 15.95& 102.32 ± 16.88 \\
        $ S_{90} \times 10^{-12} $ (Data)  & -0.649 ± 0.949  & -1.328±0.971 & -1.129±1.297 & -0.835 ± 0.94  \\
        $ S_{90} \times 10^{-12} $ (Sim)  & 2.073 ± 1.927   & 1.831 ± 1.91 & 1.963 ± 2.122 & 1.891 ± 1.876 \\
        \hline
    \end{tabular}
    \caption{The 1$\sigma$ uncertainties of $\theta_{W}$, $\theta_{H}$, and $S_{90}$ for  the SMICA map masked by the Common, L2023, H2023 and Neargal masks.}
    \label{tab:cross_points}
\end{table}\\
The p-values shown in Table~\ref{tab:pvalue_all} reflect the probability of observing the values further on the tail compared to those of the synthetic maps, obtained under the assumption of $\Lambda$CDM. As we can see, with respect to the statistics used, $\theta_H$ is the one with the lowest variation, while when we analyze the addition of the masks, the p-values of the neargal mask slightly increase for all three statistics, numerically we have 7.0$\%$ for $\theta_{W}$, 3.0$\%$ for $\theta_{H}$, and 9.0$\%$ for $S_{90}$. These results suggest that the application of the masks do not change significantly the discrepancy between the observed data and the predictions of the $\Lambda$CDM model. 
\begin{table}[h]
    \centering
    \renewcommand{\arraystretch}{1.2}
    \begin{tabular}{lcccc}
        \hline
        \textbf{ } & \textbf{Common mask} & \textbf{L2023} & \textbf{H2023} & \textbf{Neargal Mask}  \\
        \hline
        $ \theta_W $  & 1.0 $\%$  & 3.4 $\%$  & 3.4 $\%$ & 7.0$\%$   \\

        $ \theta_H $ & 1.0$\%$ & 1.0$\%$  &   0.9 $\%$ &3.0$\%$    \\

        $ S_{90}  $  & 8.0$\%$ & 3.7$\%$ & 6.6$\%$  & 9.0$\%$  \\

        \hline
    \end{tabular}
    \caption{P-values for $\theta_W$, $\theta_H$, and $S_{90}$ for different foreground masks.}
    \label{tab:pvalue_all}
\end{table}
Finally, we compute the chi-squared ($\chi^2$) statistic using Eq.~\ref{eq1}. The reduced chi-squared values for $W_{\eta}$ and $H(\theta)$ are presented in Table ~\ref{tab:chi_squared}. The application of the Neargal Mask slightly lowers the reduced chi-squared values for both $W_{\eta}$ and $H(\theta)$ compared to using the common mask alone.
\begin{table}[h]
    \centering
    \renewcommand{\arraystretch}{1.2}
    %\resizebox{\textwidth}{!}{  % Auto-resize the table to page width
    \begin{tabular}{lccccc}
        \hline
        & \textbf{$N_{JK}$} & \textbf{Common Mask} & \textbf{L2023} & \textbf{H2023} & \textbf{Neargal Mask} \\
        \hline
        $W_{\eta}$ & 128 & 6.96 (167.05 / 24) & 9.56 (248.55 / 26) & 8.92 (258.61 / 29) & 6.30 (151.09 / 24) \\
        $H(\theta)$ & 128 & 7.27 (159.85 / 22) & 9.62 (240.62 / 25) & 9.54 (257.67 / 27) & 6.92 (152.27 / 22) \\
        $W_{\eta}$ & 256 & 8.16 (252.94 / 31) & 10.06 (341.88 / 34) & 10.00 (370.13 / 37) & 8.11 (251.31 / 31) \\
        $H(\theta)$ & 256 & 8.28 (223.60 / 27) & 11.32 (339.56 / 30) & 10.55 (348.15 / 33) & 8.23 (222.27 / 27) \\
        \hline
    \end{tabular}
    %}
    \caption{Reduced chi-squared values for $W_\eta$ and $H(\theta)$ under different foreground masks and jackknife configurations. The values in parentheses indicate the total $\chi^2$ divided by the degrees of freedom (d.o.f.).}
    \label{tab:chi_squared}
\end{table}
Given that none of the foreground masks applied change significantly the CMB statistical estimators analyzed previously, in what follows we only consider the Neargal Mask to assess the effect of the new extragalactic foreground with the removal of the low-$\ell$ multipoles.

\newpage
\subsection{Impact of combining the removal of the low-$\ell$ multipoles with the Neargal Mask on CMB Statistical Estimators}
\begin{figure}[ht]
    \centering
    \begin{minipage}[b]{0.49\textwidth}
        \includegraphics[width=\textwidth]{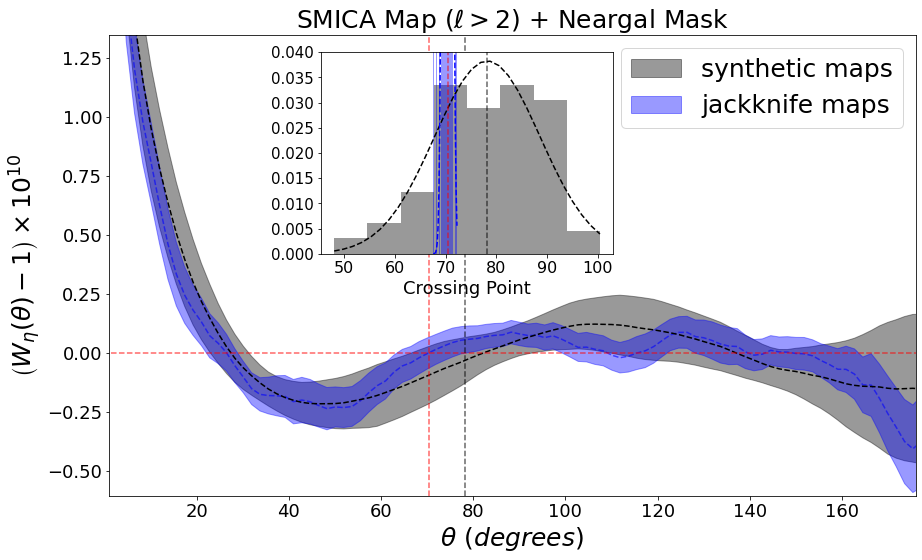}
    \end{minipage}
    \hfill
    \begin{minipage}[b]{0.49\textwidth}
        \includegraphics[width=\textwidth]{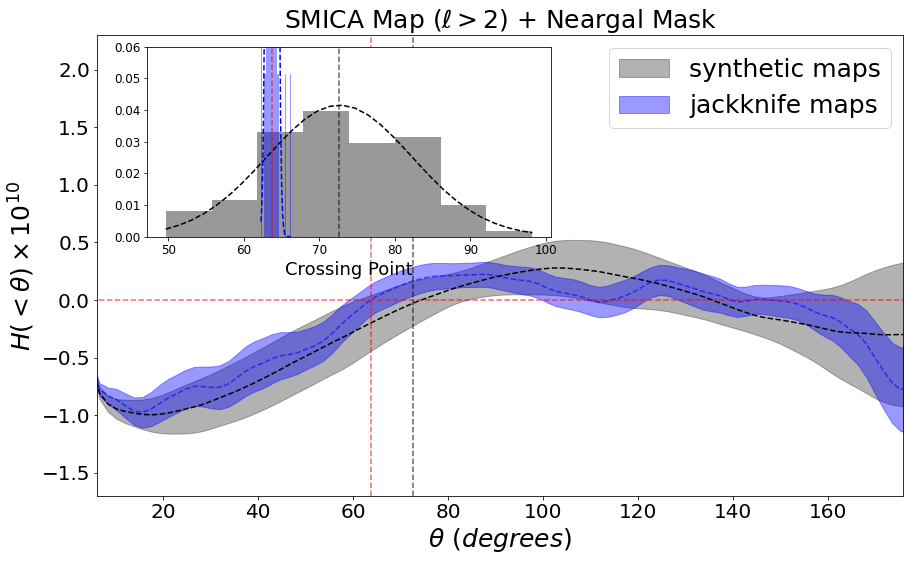}
    \end{minipage}
        \hfill
    \begin{minipage}[b]{0.49\textwidth}
        \includegraphics[width=\textwidth]{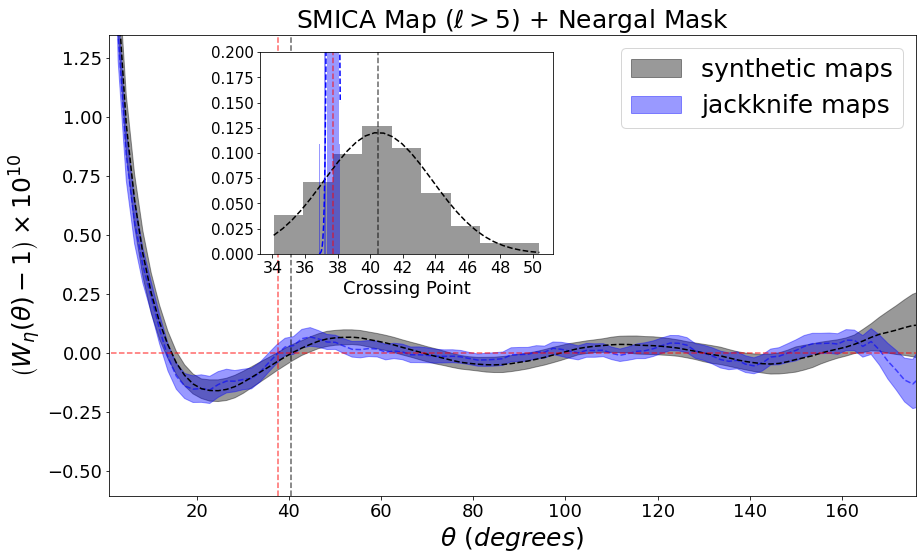}
    \end{minipage}
    \hfill
    \begin{minipage}[b]{0.49\textwidth}
   \includegraphics[width=\textwidth]{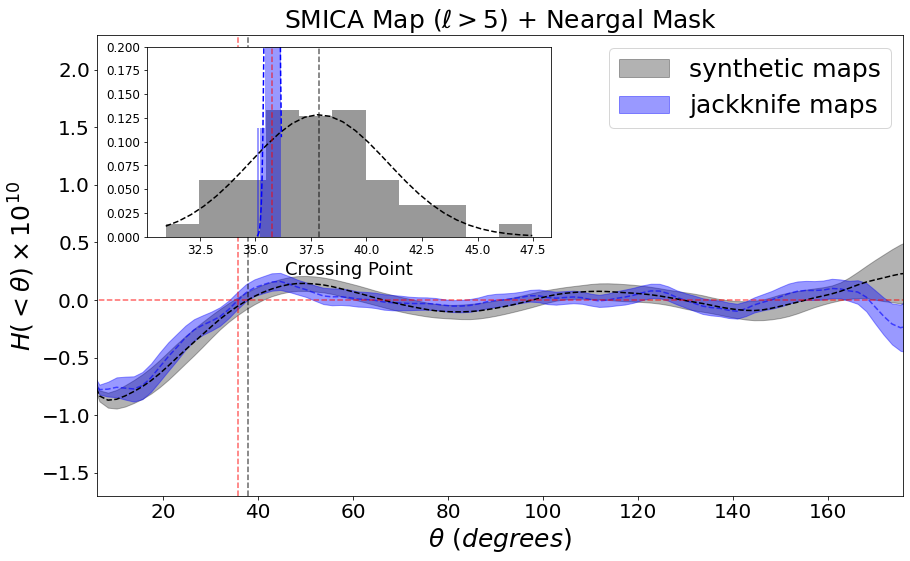}
    \end{minipage}

        \caption{The $1\sigma$ uncertainties of $W_\eta(\theta)$ (left column) and $H(\theta)$ (right column), based on  Jackknife maps and synthetic $\Lambda$CDM simulations. Top row: SMICA map with the removal of the first three multipoles; bottom row: SMICA map with the removal of the first six multipoles. In each panel, the blue dashed lines represent the SMICA data, and the blue shaded regions show the uncertainties. The black dashed lines denote the mean values from $\Lambda$CDM simulations, with the grey shaded regions representing their 1$\sigma$ dispersion. Insets display the distribution of the crossing point angles, defined where $W_\eta(\theta) = 1$ or $H(<\theta) = 0$. Vertical dashed lines in the insets mark the mean values from the SMICA data (red) and the simulations (black).  }
        \label{fig:t_wh}
\end{figure}

\begin{figure*}[htbp] %[!t]
	\centering	\includegraphics[width=0.9\textwidth]{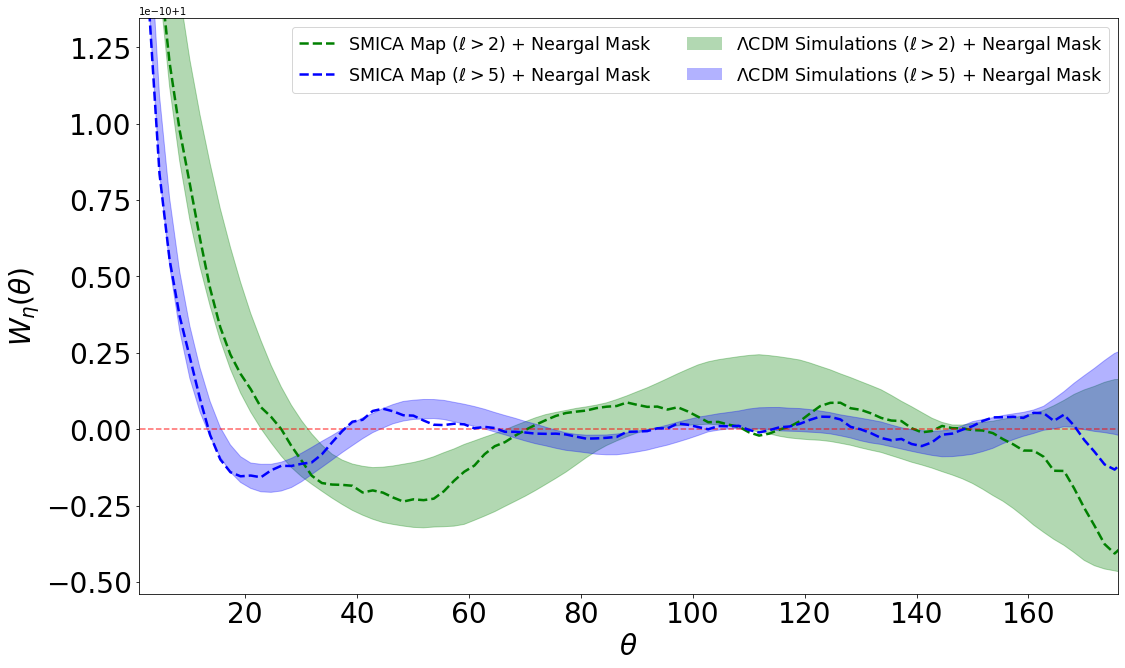}
        \caption{Angular correlation $W_{\eta}(\theta)$ for SMICA and $\Lambda$CDM maps with Neargal mask after the removal of low multipoles. The green and blue dashed lines correspond to the SMICA map with multipoles $\ell > 2$ and $\ell > 5$ retained, respectively, while the shaded regions represent the $1\sigma$ dispersion from 1000 $\Lambda$CDM simulations under the same conditions. Specifically, the green band represents simulations with $\ell > 2$ and the blue band represents $\ell > 5$, both masked by the Neargal foreground mask. }
        \label{t_whall2}
\end{figure*}

In this section, we examine the combined effects of the removal of the low-$\ell$ multipoles and the application of the Neargal Mask on the CMB statistical estimators. \\
In Figure~\ref{fig:t_wh} we present the correlation function $W_\eta(\theta)$ (left column) and homogeneity index $H(\theta)$ (right column) for the two cases, the removal of multipoles $\ell \leq 2$, and the removal of $\ell \leq 5$, respectively. We also provide the results of synthetic maps of the $\Lambda$CDM model (grey shaded regions). The data uncertainty is estimated through 128 Jackknife measurements (blue shaded regions). We notice that the removal of the quadrupole ($\ell=2$) is sufficient to provide satisfactory agreement between the $\Lambda$CDM and the CMB in both estimators $W_\eta(\theta)$ and $H(\theta)$.
In Table~\ref{tab:t_cross_points} we compare the statistical estimators $\theta_{W}$, $\theta_{H}$, and $S_{90}$ for the masked CMB maps with the removal of low multipoles, and as a consequence, the results are in great agreement with the values of $\theta_{W}$ and $\theta_{H}$ with respect to the $\Lambda$CDM predictions.
\begin{table}[h]
    \centering
    \renewcommand{\arraystretch}{1.2}
    %\resizebox{\textwidth}{!}{  % Auto-scale to page width
    \begin{tabular}{lcc}
        \hline
        & \textbf{SMICA $(\ell > 2)$ + Neargal Mask} & \textbf{SMICA $(\ell > 5)$ + Neargal Mask} \\
        \hline
        $\theta_W$ (Data) & 70.36 ± 6.78 & 37.72 ± 2.21 \\
        $\theta_W$ (Sim) & 78.20 ± 10.47 & 40.50 ± 3.33 \\
        $\theta_H$ (Data) & 63.72 ± 5.08 & 35.75 ± 1.88 \\
        $\theta_H$ (Sim) & 72.65 ± 9.66 & 37.86 ± 3.13 \\
        $S_{90} \times 10^{-12}$ (Data) & -0.776 ± 0.833 & 0.468 ± 0.569 \\
        $S_{90} \times 10^{-12}$ (Sim) & 0.930 ± 1.681 & 0.727 ± 0.887 \\
        \hline
    \end{tabular}
    %}
    \caption{The $1\sigma$ uncertainties of $\theta_W$, $\theta_H$, and $S_{90}$ for the SMICA maps after the removal of the first three and six multipoles. All maps are masked with the common mask and the Neargal mask.}
    \label{tab:t_cross_points}
\end{table}
In Figure~\ref{t_whall2} we present a comparison of the angular statistical estimators $ W_{\eta}(\theta)$ from the observations and the simulations. The results show that the extragalactic foreground masking does not affect the large impact of both quadrupole and $\ell \leq 5$ multipole subtraction. This emphasizes the significance of carefully treating large angular scales modes and foreground contamination in CMB anisotropy analyses.
 \\
In Table~\ref{tab:t_pvalue_all} we present the p-values obtained for $\theta_{W}$, $\theta_{H}$, and $S_{90}$  from the CMB maps and the $\Lambda$CDM simulations. As shown in this table, all the p-values are significantly larger than those obtained with all multipoles included.
\begin{table}[h]
    \centering
    \renewcommand{\arraystretch}{1.2}
    \begin{tabular}{lccc}
        \hline
        \textbf{ } & \textbf{SMICA $(\ell > 2)$ + Neargal Mask} & \textbf{SMICA $(\ell > 5)$ + Neargal Mask}  \\
        \hline
        $ \theta_W $  & 21.0 $\%$  & 20.0$\%$   \\

        $ \theta_H $ & 19.0$\%$  & 25.0$\%$    \\

        $ S_{90}  $  & 16.0$\%$  & 36.0$\%$  \\

        \hline
    \end{tabular}
    \caption{P-values for $\theta_{W}$, $\theta_{H}$, and $S_{90}$ derived from SMICA maps after the removal of the first three and six low multipoles.
 }
    \label{tab:t_pvalue_all}
\end{table}\\
Finally, we evaluate the reduced chi-squared ($\chi^2_{\nu}$) values for the correlation function $W_{\eta}(\theta)$ and the homogeneity index $H(\theta)$, considering the removal of  low-$\ell$ multipoles. The results are reported in Table~\ref{tab:t_combined_chi_squared}. Consistently with the p-values, the $\chi^2_{\nu}$ are close to the unity for all cases regardless the inclusion of the extragalactic foreground mask.
\begin{table}[htbp]
    \centering
    \renewcommand{\arraystretch}{1.2}
    %\resizebox{\textwidth}{!}{  % Scale to fit page width
    \begin{tabular}{lccc}
        \hline
        & \textbf{$N_{JK}$} & \textbf{SMICA $(\ell > 2)$ + Neargal Mask} & \textbf{SMICA $(\ell > 5)$ + Neargal Mask} \\
        \hline
        $W_{\eta}(\theta)$ & 128 & 1.16 (27.93 / 24) & 0.98 (37.35 / 38) \\
        $H(\theta)$        & 128 & 1.46 (32.02 / 22) & 1.05 (33.55 / 32) \\
        $W_{\eta}(\theta)$ & 256 & 1.49 (47.80 / 32) & 1.01 (46.36 / 46) \\
        $H(\theta)$        & 256 & 1.64 (47.52 / 29) & 0.95 (35.10 / 37) \\
        \hline
    \end{tabular}
    %}
    \caption{Reduced chi-squared values for SMICA maps after the removal of the first three and six multipoles,  using 128 and 256 jackknife maps. }
    \label{tab:t_combined_chi_squared}
\end{table}

\begin{figure}
    \centering \includegraphics[width=0.8\textwidth]{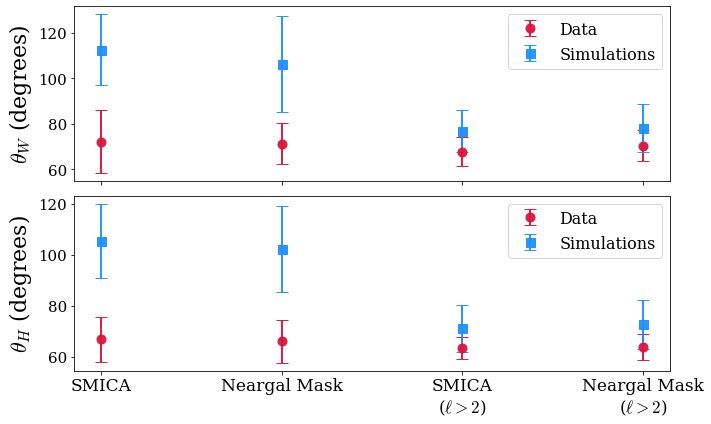}
    \caption{Comparison of $\theta_{W}$ (top) and $\theta_{H}$ (bottom) values for the SMICA map and its corresponding $\Lambda$CDM simulations, SMICA masked with the Neargal Mask and its simulations counterpart, and both scenarios after the removal of the first three multipoles.}
    \label{com_theta_h_w}
\end{figure}
\section{Conclusions}\label{sec:CONCLUSIONS}

In this paper, we have systematically assessed various methods to explore the possible origins of the discrepancies between the homogeneity scale observed in the Cosmic Microwave Background data and the theoretical predictions of the $\Lambda$CDM model. In order to do that we analyzed different quantities, as the homogeneity scale, the homogeneity index and the slope of the homogeneity index at $\theta = 90^\circ$. We performed the statistical analyses with different masks and found that masking extragalactic foregrounds using the L2023, H2023, and Neargal Mask can slightly improve the consistency between observations and theoretical expectations. However, it does not solve the observed discrepancies. \\
Our analysis shows that removing low multipoles, especially the quadrupole, from both the data and the models, significantly improves their mutual consistency. We argue that this is a clear indication that the observed low value of the CMB quadrupole contributes notably to the observed anomalies in the homogeneity scale. \\
We also added masks associated with the new extragalactic foreground to the full and the low-$\ell$ removed CMB maps. Our analysis shows similar results to those of the full CMB SMICA map, indicating that the foreground masking gives consistent results concerning the discrepancies with the $\Lambda$CDM model. Fig.~\ref{com_theta_h_w} demonstrates that the values of $\theta_{W}$ and $\theta_{H}$, both with and without foreground masking, converge more closely to the $\Lambda$CDM predictions when the first three multipoles are excluded.
Since the quadrupole of the actual data is very small, removing it does not significantly change anything in the data. What changes are the simulations which do have considerable quadrupoles. The problem of the homogeniety scale is resolved since, when removing the quadrupole from the simulations, the homogeneity scale of the simulations are shifted towards that of the real data and not the opposite.
This result shows that the observed low homogeneity scale value is associated with the quadrupole anomaly, regardless of the reported extragalactic foreground associated with nearby galaxies.

\begin{acknowledgments}
XS is supported by the Coordena\c{c}\~ao de Aperfei\c{c}oamento de Pessoal de N\'ivel Superior (CAPES). CB acknowledged financial support from Funda\c{c}\~ao de Amparo \`a Pesquisa do Estado do Rio de Janeiro (FAPERJ). RSG thanks financial support from FAPERJ grant No. 260003/005977/2024 - APQ1. JSA is supported by CNPq grant No. 307683/2022-2 and FAPERJ grant No. 259610 (2021). This work was developed thanks to the use of the National Observatory Data Center (CPDON).  Some of the results were derived using the HEALPix package\cite{Healpix.2005ApJ...622..759G}.
\end{acknowledgments}

\printbibliography

\appendix
\section{Multipole removal}
\begin{figure}[ht]
    \centering
    \includegraphics[width=\textwidth]{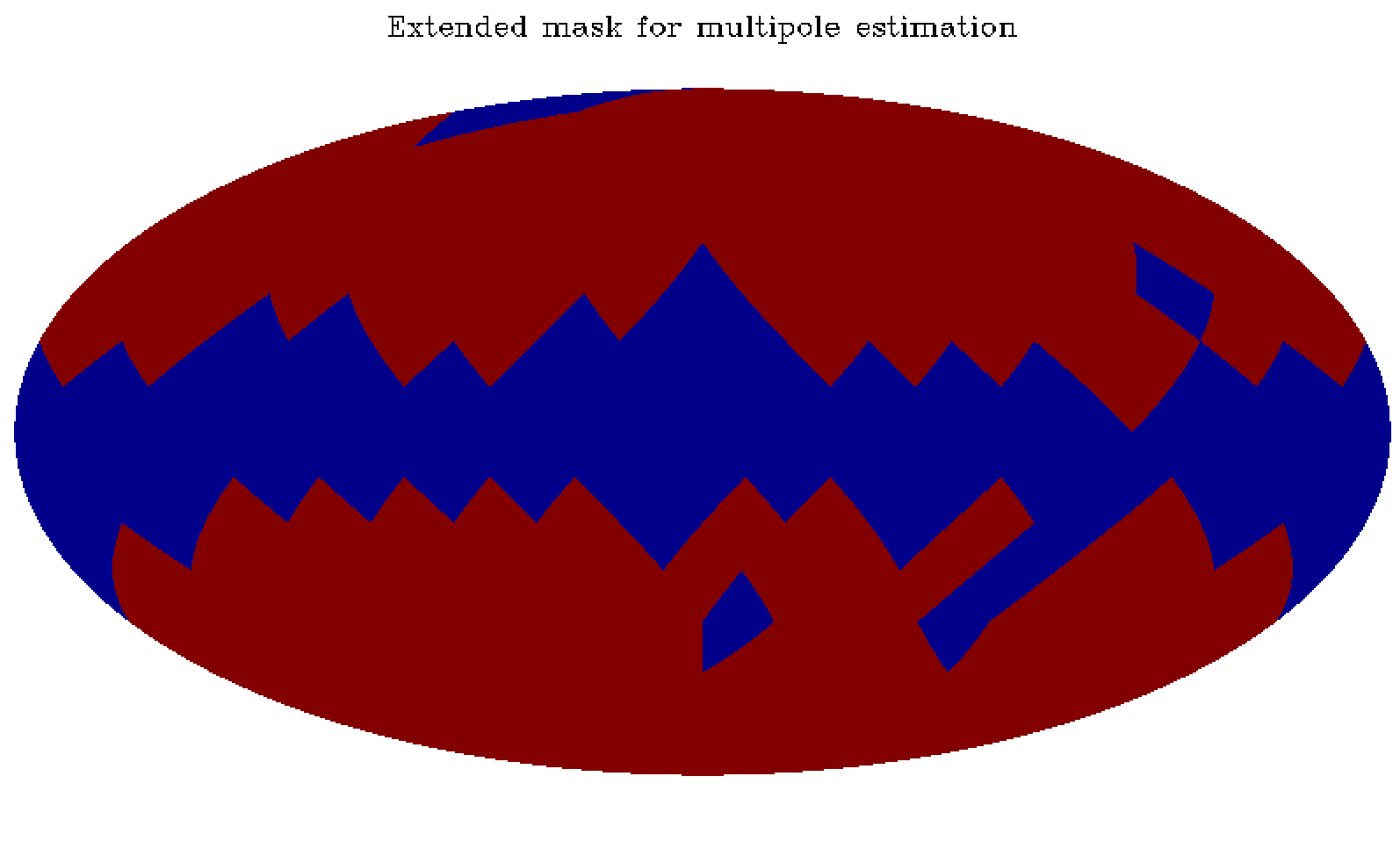}
    \includegraphics[width=\textwidth]{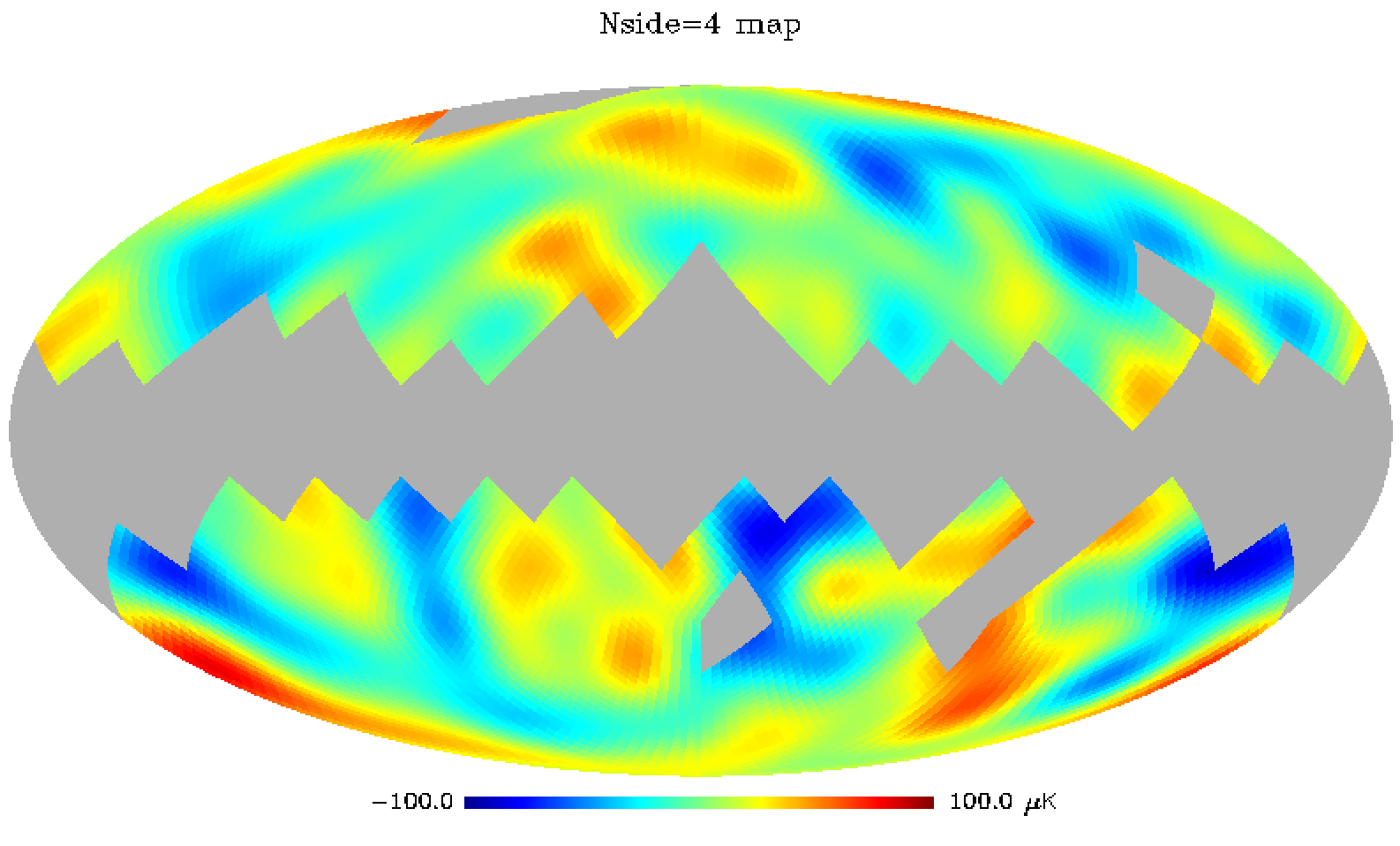}
    \caption{The extended mask and the SMICA map reduced to Healpix resolution $N_\mathrm{side} = 4$ used for estimating low multipole modes.\label{fig:appendix}}
\end{figure}
Figure \ref{fig:appendix} shows the extended mask and the $N_\mathrm{side} = 4$  map used for estimating individual multipole moments. All multipole maps up to $\ell=\mathrm{16}$ are estimated and the map reconstructed from the estimated multipole moments equals the input map at numerical precision. The first multipole moments are used in the main text to remove the largest scales before obtaining $W_\eta$.

\end{document}